\documentclass[10pt,a4paper,english,twocolumn]{IEEEtran} 
\usepackage[T1]{fontenc}
\usepackage{color}
\usepackage{babel}
\usepackage{verbatim}
\usepackage{textcomp}
\usepackage{amsmath}
\usepackage{amstext}
\usepackage{graphicx}
\usepackage{caption}
\usepackage{subcaption}
\usepackage[unicode=true,pdfusetitle,
 bookmarks=false,
 breaklinks=false,pdfborder={0 0 1},backref=false,colorlinks=true]
 {hyperref}
\hypersetup{
 linkcolor=blue, citecolor=red}

\makeatletter

\pdfpageheight\paperheight
\pdfpagewidth\paperwidth

\providecommand{\tabularnewline}{\\}

\usepackage{epstopdf}
\usepackage{cite}
\usepackage{citesort}
\usepackage{balance}

\makeatother

\begin{document}

\title{A Brief History on the Theoretical Analysis of Dense Small Cell Wireless Networks
}

\author{
{David L$\acute{\textrm{o}}$pez-P$\acute{\textrm{e}}$rez}$^{\dagger},$ \textit{IEEE Senior Member},
\thanks{$^{\dagger}$David L$\acute{\textrm{o}}$pez-P$\acute{\textrm{e}}$rez is with Nokia Bell Labs, Ireland (email: david.lopez-perez@nokia-bell-labs.com)
$-$ ``\emph{I would like to dedicate this article to my beloved nephew, Lucas,
who was born the day I had the strength to finish the writing of the draft of this article. 
As the first sunlight rays of spring, he provided me the inspiration}''.}
{Ming Ding}$^{\ddagger},$ \textit{IEEE Senior Member}
\thanks{$^{\ddagger}$Ming Ding is with Data61, CSIRO, Australia (e-mail: Ming.Ding@data61.csiro.au)
$-$ ``\emph{I would like to dedicate this article to my brilliant wife, Minwen Zhou, who constantly inspires me during my scientific journey and in all aspects of my life. She also came up with the name of "network capacity crawl/crash" in March 2016, which vividly describes the caveats of wireless network densification to be explored in this article.}''.}
}
\maketitle
\begin{abstract}

This article provides dives into the fundamentals of dense and ultra-dense small cell wireless networks, 
discussing the reasons why dense and ultra-dense small cell networks are fundamentally different from sparse ones,
and why the well-known linear scaling law of capacity with the base station (BS) density in the latter does not apply to the former.
In more detail, 
we review the impact of the following factors on ultra-dense networks (UDNs), 
\emph{(i)} closed-access operations and line-of-sight conditions,
\emph{(ii)} the near-field effect, 
\emph{(iii)} the antenna height difference between small cell BSs and user equipments (UEs), 
and \emph{(iv)} the surplus of idle-mode-enabled small cell BSs with respect to UEs.
Combining all these network characteristics and features,
we present a more realistic capacity scaling law for UDNs,
which indicates \emph{(i)} the existence of an optimum BS density to maximise the area spectral efficiency (ASE) for a given finite UE density, 
and \emph{(ii)} the existence of an optimum density of UEs that can be simultaneously scheduled across the network to maximise the ASE for a given finite BS density.

\end{abstract}


\section{Introduction\label{sec:Introduction}}

Dense and massive,
meaning more of everything,
are the key adjectives that will best describe the fifth generation (5G) and the future of cellular technology.
Supported by the recent advancements in radio frequency equipment as well as digital processing capabilities,
the rationale behind this philosophy of deploying more network infrastructure follows from the renowned \emph{Shannon-Hartley theorem}~\cite{1948Shannon},
i.e.,~$C{\rm[bps]} = B{\rm[Hz]}\log_{2}(1+\gamma{\rm[\cdot]})$.
This theorem,
probably the most important result in the field of information theory,
indicates that the end-user capacity,~$C$ (in $bps$),  
\begin{enumerate}
\item
increases linearly with the amount of spectrum,~$B$ (in Hz), that each device can access, but 
\item
only increases logarithmically with its signal quality, represented by its signal-to-interference-plus-noise ratio (SINR),~$\gamma$, which is a scalar.
\end{enumerate}
This suggests that the best approach to enhance the end-user capacity is to maximise the amount of spectrum that the end-user can access,
while providing it a reasonably good SINR.
Due to the logarithmic scaling law with respect to the signal quality, 
it is clear that investing in enhancing SINR loses momentum after some point.

To realise such maximisation of spectrum per end-user,
it is necessary that networks move away from the current cell-centric paradigm to a user-centric one.
In other words,
operators should abandon the current paradigm,
where a large number of user equipments (UEs) share the radio resources operated by one base station (BS),
and embrace a new network,
in which all radio resources can be made available anytime and anywhere to a given UE.
This will create the feeling of an infinite, or at least, a very large network capacity at the UE.

Two main technologies are currently driving such a paradigm shift in the telecommunication industry:
ultra-dense networks (UDNs)~\cite{Andrews2014}~\cite{Tutor_smallcell} and massive multiple-input multiple-output (mMIMO)~\cite{6415388}~\cite{2014Marzetta}.
Although different in nature,
and facing a diverse set of technical and business challenges, 
both frameworks aim to achieve the same goal,
i.e., \emph{reaching an operating point where there is one-UE-per-cell or one-UE-per-beam, respectively, 
with each UE thus reusing the entire available spectrum}.
As the available bandwidth foreseen in the 5G spectrum is large,
from 400MHz in the sub-6GHz bands to 2GHz in the new mmWave bands,
these paradigms can lead to a tremendous throughput of more than 1Gbps per UE on average~\cite{Tutor_smallcell}~\cite{6736746}~\cite{6824752}.
Even larger throughputs are attainable,
if UEs are equipped with more than one antenna,
and are able to receive multiple layers of spatially multiplexed data~\cite{917094}~\cite{1197843}.

In this article,
we walk the readers through the exciting history of sub-6GHz UDNs,
a feature that will be of significant importance in 5G networks~\cite{Andrews2014}~\cite{Tutor_smallcell} and beyond.
In more detail, we focus on its theoretical modelling,
and show how researchers have stood on `the shoulders of giants' to develop the current theory of dense networks.
Moreover,
we present the latest developments on the understanding of UDNs and their capacity scaling laws,
which we believe are the most accurate up to date.

The rest of this article is structured as follows: 
In Section~\ref{sec:history},
the traditional theoretical understanding on network densification and capacity scaling up to 2012 is introduced, 
which will be used as a benchmark in the rest of this tutorial article. 
In Section~\ref{sec:access}, 
the effect of closed- versus open-access operation in such traditional theoretical understanding is presented.
In Section~\ref{sec:probLoS}, 
we discuss on the impact of line-of-sight (LoS) transmissions in UDNs,
a game changer. 
In Section~\ref{sec:nearFieldEffect} and Section~\ref{sec:antennaHeight}, 
we further elaborate on the impact of the near-field effect and the UE-BS antenna heights difference (a topic under careful scrutiny nowadays), respectively.
Combining all these ingredients and considering idle-mode-enabled BSs,
a more realistic definition and capacity scaling law for UDNs is put forward in Section~\ref{sec:IMC_and_newLaw}.
Finally, the conclusions are drawn in Section~\ref{sec:Conclusion}.

\section{The Old Understanding \label{sec:history}}

It is hard to point out who came up with the idea of the small cell BS,
as the concepts of microcell and picocell BSs have been around for many years now.
However,
one can think of the base station router (BSR)~\cite{Bauer2007},
developed by Bell Laboratories around the year 2007,
as the father of today's small cell BS;
the true enabler of cellular network densification,
which successfully took the idea into the market for the first time.

A large amount of research and engineering work were behind the success of the small cell BS.
Remarkable is the work of H. Claussen \emph{et. al.} on small cell auto-configuration and self-optimisation in the early days of the concept~\cite{2008Claussen}~\cite{4547576}~\cite{5199033},
as well as the effort of the Small Cell Forum~\cite{smallCellForum} and the Third Generation Partnership Project (3GPP)~\cite{3gpp}  to promote and standardise this technology up to now.

When it comes to small cell deployment and network wide performance,
the theoretical work of M. Haenggi, J. G. Andrews, F. Baccelli, O. Dousse and M. Franceschetti stands out.
In their seminal work~\cite{Haenggi2009}~\cite{Jeffery2011TWC_coverage},
the authors created a mathematical framework based on stochastic geometry to analyse the performance of random networks in a tractable manner.

In a nutshell, 
this mathematical stochastic geometry framework allows to theoretically calculate, 
sometimes even in a closed-form, 
the typical UE's SINR coverage probability,
which is defined as the probability that the typical UE's SINR, $\gamma$, is larger than a threshold, $\gamma_0$,
i.e., $\textrm{Pr}\left[\mathrm{\gamma}>\gamma_0\right]$.
Based on this coverage probability,
also known as success probability,
the SINR-dependent area spectral efficiency (ASE) in $\textrm{bps/Hz/km}^{2}$ can also be investigated,
among other metrics.
This framework has become the \emph{de facto} tool for the theoretical analysis of small cell networks in the entire wireless community.
Good tutorials and more references on the fundamentals of this mathematical framework can be found in~\cite{Baccelli2009},~\cite{Haenggi2012},~\cite{ElSawy2013},~\cite{Mukherjee2014} and the references therein.

Many efforts have been made since 2009 to extend the capabilities of this stochastic geometry framework for enhancing the understanding of small cell networks. 
M. Haenggi \emph{et. al.} further developed the framework to account for 
different stochastic processes~\cite{6841639}, 
and distinct performance metrics, 
such as the typicality of the typical user~\cite{Haenggi2016TWC_meta} and the transmission delay~\cite{6353585}, 
among others. 
T. D. Novlan \emph{et. al.} further extended the framework to study uplink transmissions, 
calculating the aggregated interference using the probability generating functional (PGFL) of the homogeneous Poisson point process (PPP)~\cite{Novlan2013_TWC_uplink}.
M. Di Renzo \emph{et. al.} brought it to a new level, 
by considering more detailed wireless channel characteristics in the modelling, 
such as non-HPPP distributions, building obstructions, shadowing, and non-Rayleigh multi-path fading, of course, at the expense of tractability~\cite{7482733}. 
%
When it comes to the analysis of different wireless network technologies and features using stochastic geometry,
it is worth highlighting the extensive work of J. G. Andrews, V. Chandrasekar, H. S. Dhillon \emph{et. al.},
which touches on  
spectrum allocation~\cite{5288507}, 
sectorisation~\cite{5165314}, 
power control~\cite{5200991},
MIMO~\cite{6596082},
small cell-only networks~\cite{Jeff2011},
multi-tier heterogeneous networks~\cite{Dhillon2012hetNetSG}, 
load-balancing~\cite{6463498}, 
device-to-device communications~\cite{Chun2017TWC_D2D_mu_fading}, 
content caching~\cite{MalakTWC2018_contentcachingD2D},
Internet of Things (IoT) networks~\cite{Kouzayha2018TWC_iot}, 
and unmanned aerial vehicles~\cite{Chetlur2017Tcom_UAV}, 
to cite a few. 
When about the massive use of antennas and higher frequency bands,
the studies of R. W. Heath, T. Bai \emph{et. al.} stand out,
for example, 
those on 
mMIMO~\cite{6894455} and mmWave performance analysis~\cite{6932503},
random blockage~\cite{Gupta2018TWCrandomBlockages}, 
mmWave ad-hoc networks~\cite{Thornburg2018TWC_mmwave_adhoc}, 
mmWave secure communications~\cite{Yongxu2017TWC_secure_MMwave},
shared mmWave spectrum~\cite{Jurdi2018TCCN_mmwave}~\cite{Park2018TCCN_BScoodinationMMwave}, 
and wireless power systems~\cite{Lifeng2017JSAC_powereddensenetwork}.
For further reference, 
and with regard to the analysis of other relevant network aspects, 
it is worth pointing out the studies  of 
G. Nigam \emph{et. al.} on coordinated multipoint joint transmission~\cite{6928420},
H. Sun \emph{et. al.} on dynamic time division duplex~\cite{Hongguang2016_JSAC_DynamicTdd},
and Y. S. Soh \emph{et. al.} on energy efficiency~\cite{6502479}. 
Many other analyses studying different types of stochastic processes, performance metrics, wireless characteristics and network features are there now. 

\bigskip

\begin{figure}
    \centering
        \includegraphics[width=0.3\textwidth]{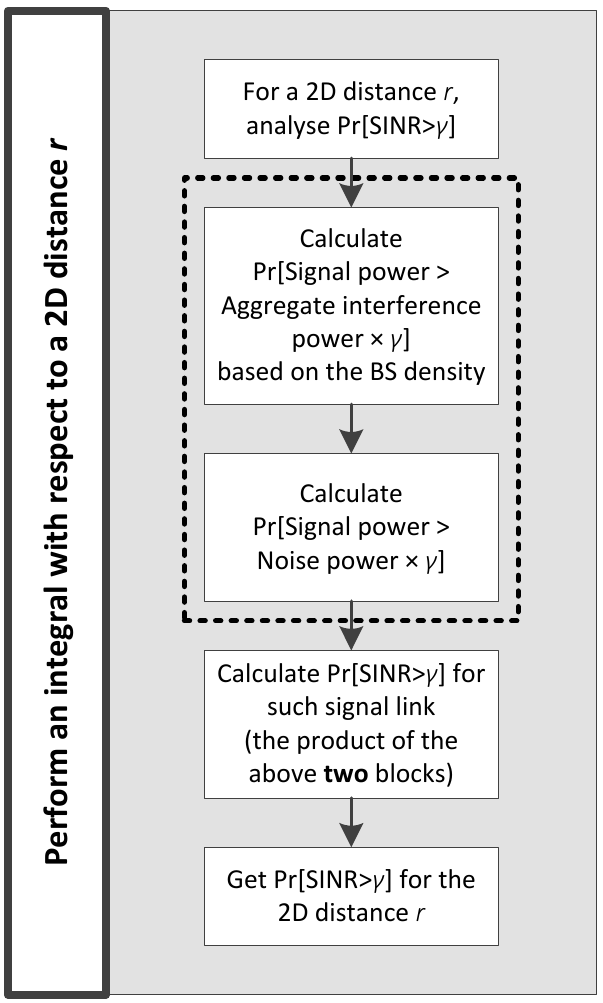}
        \caption{Logical steps within the standard stochastic geometry framework to obtain the results derived in~\cite{Jeff2011}, 
        i.e., \emph{the SINR invariance law}.}
        \label{fig:th1}
\end{figure}

\begin{figure*}
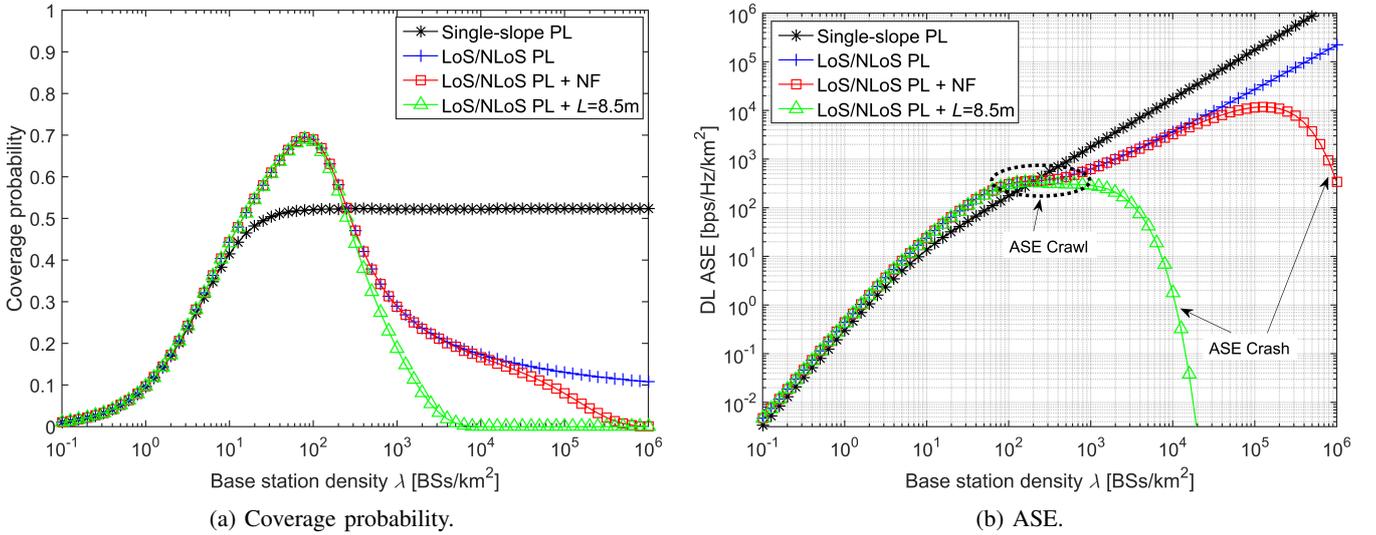

    \centering
    \begin{subfigure}[b]{0.48\textwidth}
        \includegraphics[width=\textwidth]{Mag2017_PrCov_infUE_v1}
        \caption{Coverage probability.}
        \label{fig:covProb1}
    \end{subfigure}
    \hspace{0.1cm}
    \begin{subfigure}[b]{0.48\textwidth}
        \includegraphics[width=\textwidth]{Mag2017_ASE_infUE_v1}
        \caption{ASE.}
        \label{fig:ase1}
    \end{subfigure}
    \caption{The decline in coverage probability and ASE originated due to channel and deployment issues.
    Note that all the results are obtained using practical 3GPP channel models described in Table~\ref{tab:key_param}.
    }\label{fig:theDecline}
\end{figure*}

\begin{table*}
\caption{\label{tab:key_param}Parameter values recommended by the 3GPP~\cite{TR36.828}}
\centering{}{\small{}}%
\scalebox{1}{
\begin{tabular}{|l|l|}
\hline
{\small{}Parameters} & {\small{}Assumptions}\tabularnewline
\hline
\hline
{\small{}BS and UE distribution} & {\small{}Spatial Homogeneous Poisson Point Process}\tabularnewline
\hline
{\small{}BS transmission power} & {\small{}24\,dBm (on a 10\,MHz bandwidth)}\tabularnewline
\hline
{\small{}Noise power} & {\small{}-95\,dBm (on a 10\,MHz bandwidth)}\tabularnewline
\hline
{\small{}The BS-to-UE LoS path loss} & {\small{}$103.8+20.9\log_{10}r$ ($r$ in km)}\tabularnewline
\hline
{\small{}The BS-to-UE NLoS path loss} & {\small{}$145.4+37.5\log_{10}r$ ($r$ in km)}\tabularnewline
\hline
{\small{}The BS-to-UE LoS prob. function} & {\small{}$\begin{cases}
\hspace{-0.2cm}\begin{array}{l}
1\hspace{-0.1cm}-\hspace{-0.1cm}5\exp\left(-0.156/r\right),\\
5\exp\left(-r/0.03\right),
\end{array}\hspace{-0.3cm}\hspace{-0.1cm}\hspace{-0.1cm} & \begin{array}{l}
0<r\leq0.156\,\textrm{km}\\
r>0.156\,\textrm{km}
\end{array}\end{cases}$}\tabularnewline
\hline
\end{tabular}{\small \par}}
\vspace{-0.1cm}
\end{table*}

\begin{figure*}
    \centering

    \begin{subfigure}[b]{0.9\textwidth}
        \includegraphics[width=\textwidth]{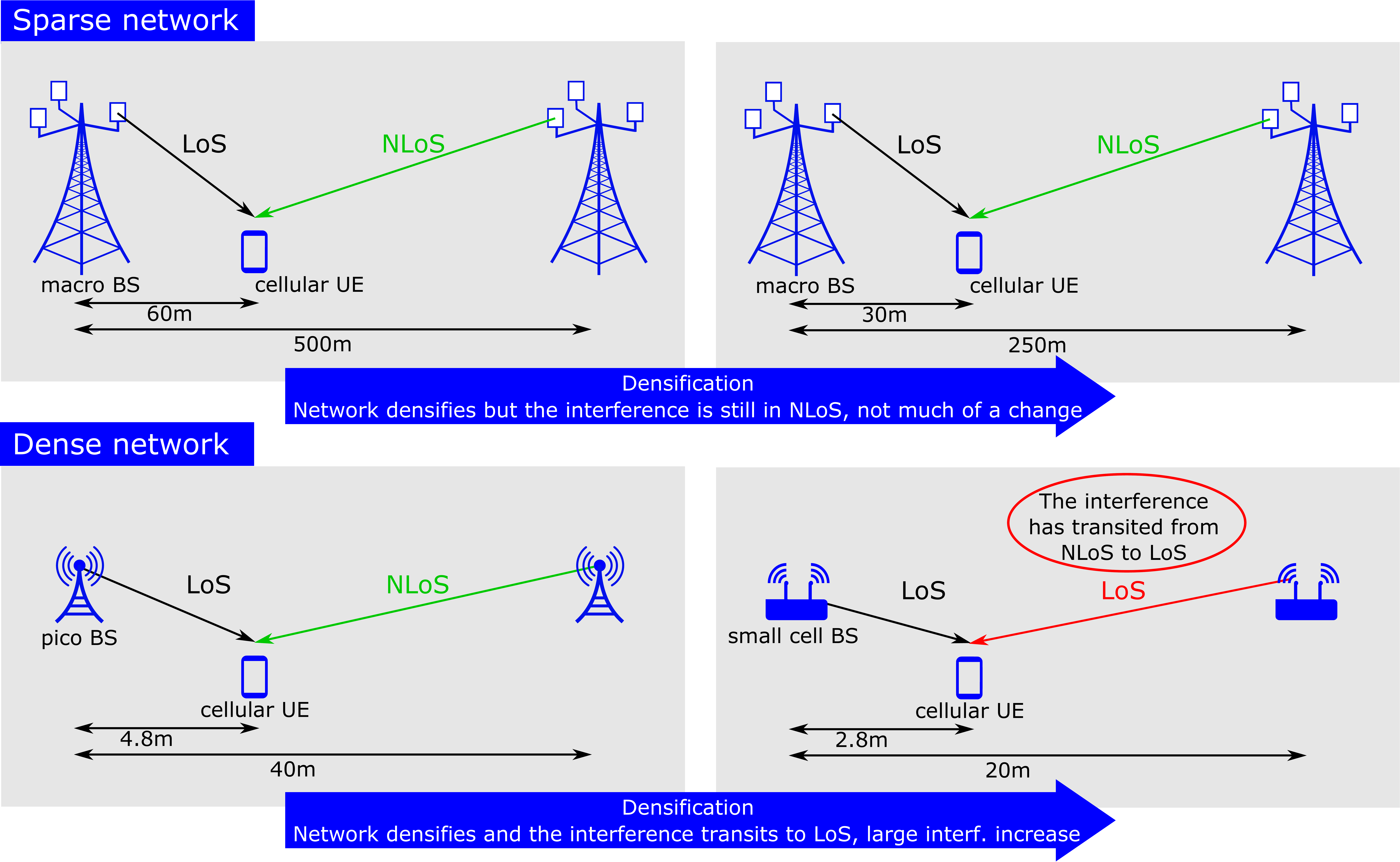}
        \caption{NLoS-to-LoS transition.}
        \label{fig:nlos2losTransition}
    \end{subfigure}
    \begin{subfigure}[b]{0.9\textwidth}
    \vspace{0.8cm}
        \includegraphics[width=\textwidth]{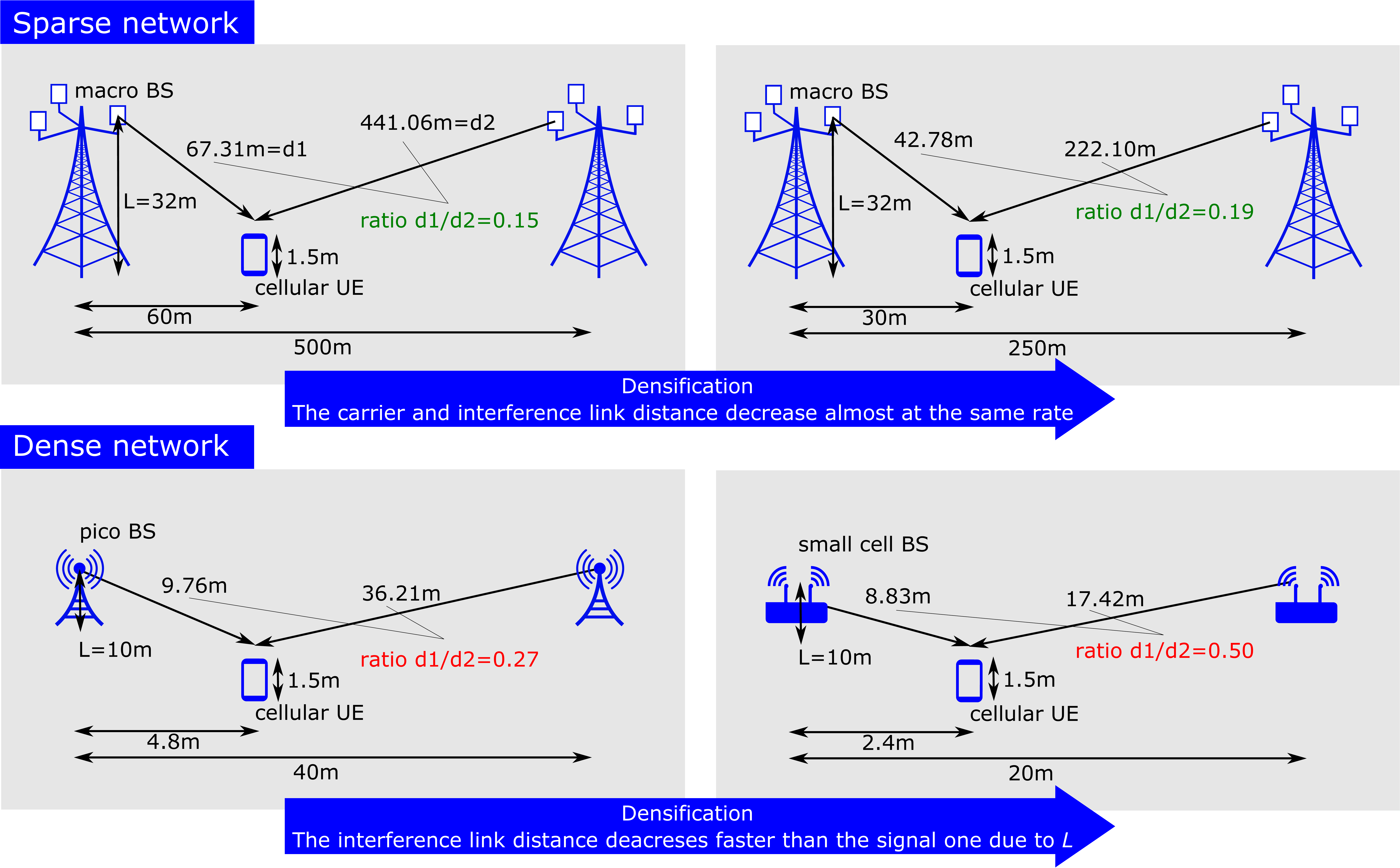}
        \caption{Antenna height difference.}
        \label{fig:antennaHeight}
    \end{subfigure}
    \caption{Reasons for signal quality decline in UDNs}\label{fig:CrashReasons}
\end{figure*}

Among all these results,
one of the most important theoretical findings is that by J. G. Andrews and H. S. Dhillon \emph{et. al.}, 
concluding that 
the fears of an interference overload in small cell networks were not well-grounded~\cite{Jeff2011}~\cite{Dhillon2012hetNetSG}.
Instead, their results showed that the increase in interference power due to the larger number of interfering BSs in a dense network is exactly counterbalanced by the increase in signal power due to the closer proximity of transmitters and receivers.
This was observed for both small cell-only networks~\cite{Jeff2011} and complex heterogeneous networks with many different classes of BSs~\cite{Dhillon2012hetNetSG}.
This conclusion is powerful,
meaning that an operator can continually densify its network,
no problem,
and expect that the spectral efficiency in each cell stays roughly constant,
or in other words,
that the network capacity linearly grows with the number of deployed cells.
For the sake of simplicity, 
we will refer to this finding as \emph{the SINR invariance law}.
The pseudo code that depicts the necessary steps to compute these results using stochastic geometry can be found in Fig.~\ref{fig:th1}.
Note that the probability of coverage is computed by the product of two probabilities as illustrated in the dashed block in Fig.~\ref{fig:th1}, i.e., 
\begin{enumerate}
\item
the probability of the UE's signal power being larger than the aggregate interference power times the threshold $\gamma$, and 
\item
the probability of the UE's signal power being larger than the noise power times the threshold $\gamma$.
\end{enumerate}

For illustration purposes,
the black curves with star markers in Fig.~\ref{fig:theDecline} illustrate such a law, 
i.e., the results in~\cite{Jeff2011},
indicating how the coverage probability stays constant and the ASE roughly grows linearly, 
as the BS density increases.
This exciting message created a big hype in the industry,
presenting the small cell BS as the ultimate mechanism in providing a superior broadband experience:
\emph{`Just deploy more cells and everything will be fine'}.
However, a few important caveats to realise such a view in a UDN were quickly found.
Among them, it is worth highlighting the need for an open-access operation~\cite{2008Claussen}~\cite{5394027}
and the impact of the transition of a large number of interfering links from non-line-of-sight (NLoS) to LoS conditions~\cite{our_GC_paper_2015_HPPP}~\cite{our_work_TWC2016}.
These two issues are tackled in the following two sections.

\section{The Effect of the Access Method \label{sec:access}}

Closed-access operation provides a Wireless Fidelity (Wi-Fi) access point like experience,
in which the owner of the small cell BS can select which UEs can associate to it.
This is an appealing business model,
but prevents the UE to connect to the strongest cell.
This degrades the UE's SINR and violates the SINR invariance law in~\cite{Jeff2011}~\cite{Dhillon2012hetNetSG}.
This is because the interference power can now grow much faster than the signal power.
This is particularly true when the UE moves away from its closed-access cell (which it can access) and gets closer to a neighbouring one (which it cannot access).
Open-access operation has been widely adopted in small cell BS products to address this issue, 
and restore the SINR invariance law~\cite{2008Claussen}~\cite{5394027}.

\section{The Impact of the NLoS to LoS Interference Transition \label{sec:probLoS}}

A more fundamental problem was found in~\cite{our_GC_paper_2015_HPPP}~\cite{our_work_TWC2016},
which shows that the interfering power can grow faster than the signal power with the consequent signal quality degradation,
even if open-access operation is adopted,
when the BS density increases towards a UDN.

To understand this phenomena,
it is important to note that the SINR invariance law presented in~\cite{Jeff2011}~\cite{Dhillon2012hetNetSG} was obtained assuming a single-slope path loss model,
i.e., both the interference and the signal power decay at the same pace $d^{-\alpha}$ over distance $d$,
where $\alpha$ is the path loss exponent.
Although simplistic,
when the path loss exponent is `fine-tuned',
this single-slope path loss model is applicable to sparse networks,
such as macrocell and microcell ones.
However,
this model is inaccurate for dense networks where the small cell BSs are deployed below the clutter of man-made structures.
This is because the probability of the signal strength abruptly changing, 
due to a change in the LoS condition of the link, 
is much larger in a dense network where the small cell BSs are deployed at the street level.

\begin{figure}[t]
        \centering
        \includegraphics[width=0.48\textwidth]{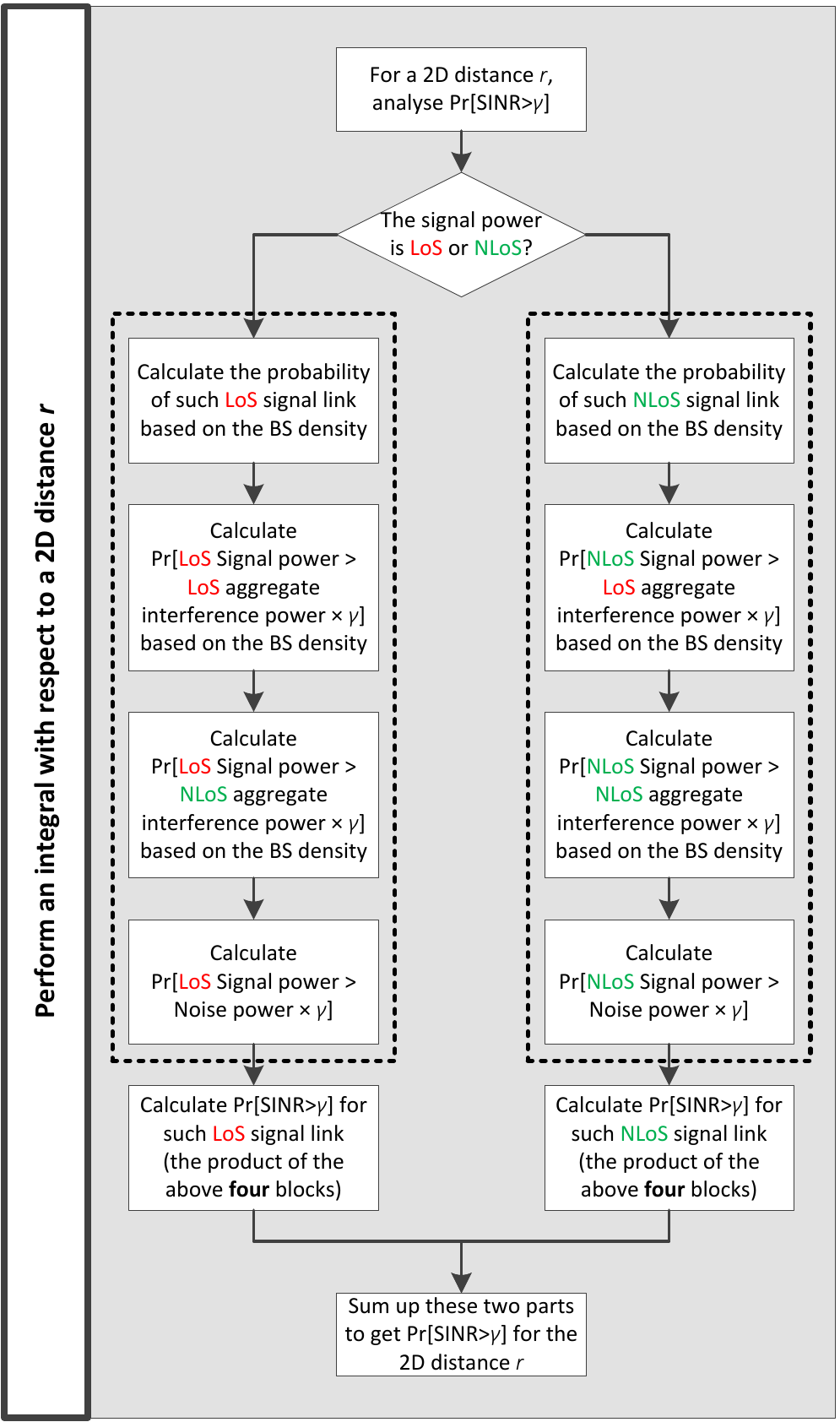}
        \caption{Logical steps within the standard stochastic geometry framework to obtain the results in~\cite{our_GC_paper_2015_HPPP}~\cite{our_work_TWC2016}, 
        considering NLoS and LoS transmissions, which yield the \emph{ASE Crawl} phenomena.}
        \label{fig:th2}
    \end{figure}

To model this critical channel characteristic,
whether the UE is in NLoS or LoS with a BS,
the authors in~\cite{our_GC_paper_2015_HPPP}~\cite{our_work_TWC2016} presented an enhancement of the analysis in~\cite{Jeff2011},
where both a 3GPP multi-slope path loss model with NLoS and LoS components
and a probabilistic function governing the switch between them were considered.
Intuitively speaking, a key difference between the single-slope and this new multi-slope path loss model is that a UE always associates to the nearest BS in the former,
while a UE may be connected with a farther but stronger BS in the latter.
This probabilistic model introduces randomness and renders decreasing distances not as useful. 
The results of this new analysis showed an important fact:
There is a BS density region where the strongest interfering links transit from NLoS to LoS conditions,
while the signal ones stay LoS dominated due to the close proximity between the UEs and their serving BSs.
Fig.~\ref{fig:nlos2losTransition} depicts this NLoS to LoS interference transition,
which degrades the UE's SINR,
as the interference power increases at a faster pace than the signal one.
As a result,
the spectral efficiency in the cell does not stay constant,
and the network capacity does not grow linearly with the BS density anymore.
The pseudo code that depicts the enhancements to the stochastic geometry framework necessary to compute these new results can be found in Fig.~\ref{fig:th2}.
On top of the logic illustrated by Fig.~\ref{fig:th1}, 
the probability of the UE's signal power being larger than the aggregate interference power times the threshold $\gamma$, 
is further broken down into LoS/NLoS signal/interference parts, 
as illustrated in the dashed blocks in Fig.~\ref{fig:th2}. 

The blue curves with plus markers in Fig.~\ref{fig:theDecline} illustrate the results in~\cite{our_work_TWC2016},
showing how the coverage probability decreases when the BS density is larger than $10^{2}$\,$\textrm{BSs/km}^{2}$,
and how the ASE does not grow linearly  in the BS density region around  $10^{2}\sim10^{3}$\,$\textrm{BSs/km}^{2}$.
We will refer to this phenomena, 
the loss of linearity in the ASE growth due to the NLoS to LoS interference transition, 
as \textbf{the ASE Crawl} here after.
It is also important to note that the results in~\cite{our_work_TWC2016} suggest a better ASE performance than those in~\cite{Jeff2011}, 
when the BS density is smaller than $10^{2}$\,$\textrm{BSs/km}^{2}$.
This is because the propagation conditions are better, i.e.. path loss exponent is smaller,  in the former than that in the latter, 
which does not differentiate among LoS and NLoS. 

The impact of such a NLoS to LoS interference transition and the accuracy of this theoretical results were confirmed by the real-world experiments in~\cite{Qualcomm2014},
in which a densification factor of 100x (from 9 to 1107\,$\textrm{BSs/km}^{2}$) led to a network capacity increase of 40x (from 16 to 1107\,$\textrm{bps/Hz/km}^{2}$),
clearly, not a linear increase.
Moreover, following theoretical analyses considering Rician and Nakagami channel models instead of a Rayleigh one have also acknowledged this phenomena~\cite{7511506}. 

This new theoretical finding showed that the BS density matters!
However, although an inconvenient fact for UDNs,
it is important to note that once the most dominant interfering links transit from NLoS to LoS,
both the interference and signal power will again grow at a similar pace,
as they are all LoS dominated.
This restores an almost linear increase of the ASE with the BS density,
as the path loss is dominated by a similar decay rate.
This is illustrated by the blue curves with plus markers in Fig.~\ref{fig:theDecline} for BS densities lager than $10^{3}$\,$\textrm{BSs/km}^{2}$.

\textbf{Takeaways:}
The need for an open-access operation and the impact of the transition of a large number of interfering links from NLoS to LoS conditions
served as a wakeup call to the theoretical research community,
which began to realise the importance of accurate network and channel modelling,
and started to review their understanding of UDNs.
Some asked themselves whether some other important details were overlooked.
Details that could change the performance trends expected for UDNs until then.
This brought back again the original question of \emph{whether the network capacity will linearly grow with the BS density}.
Two frameworks, one on the near-field effect~\cite{related_work_Jeff}~\cite{Liu2016NF} and the other one on the antenna height difference between UEs and BSs~\cite{Ding2016GC_ASECrash}~\cite{8057291}, 
should be highlighted in this quest,
which will be presented in the next two sections.

\section{The Myth of the Near-Field Effect\label{sec:nearFieldEffect}}

While looking at a more accurate channel model that could reveal new findings,
the authors in~\cite{related_work_Jeff} presented a reasonable conjecture,
indicating that the path loss exponent should be an increasing function of the distance,
and proposed to capture this in a multi-slope path loss model similar to that presented in~\cite{our_work_TWC2016}.
To illustrate their thinking,
the authors provided the following example:
\emph{"There could easily be three distinct regimes in a practical environment:
a first distance-independent "near-field" where $\alpha_1=0$;
second, a free-space like regime where $\alpha_2=2$;
and finally some heavily-attenuated regime where $\alpha_3 > 3$,"}
and then posed the following question:
\emph{"What happens if densification pushes many BSs into the near-field regime?"}.

The mathematical results derived in~\cite{related_work_Jeff} provided an answer to such a question,
and concluded that the interference power can grow faster than the signal power when the network is ultra-dense,
even if both signals are LoS dominated.
The intuition behind is that when the UE enters the near-field range,
the signal power is bounded,
as the path loss is now independent of the distance between such UE and its serving BS, $\alpha_1=0$,
while the interference power continues to grow,
since more and more interfering small cell BSs approach the UE when the network marches into a UDN.
As a result,
once the signal power enters the near-field range,
the UEs' SINRs cannot be kept constant, 
and will monotonically decrease with the BS density.
This raised the alarm again,
as this finding indicates that the near-field effect could lead to a zero throughput in an extreme densification case
due to the overwhelming interference.

Subsequent results on the topic based on measurements,
however, have shown that this alarm was unfounded~\cite{Liu2016NF}.
These measurements,
shown in Fig.~1 of~\cite{Liu2016NF},
indicate that the near-field effect only takes place at sub-meter distances in practical UDNs with a carrier frequency of around 2 GHz and an antenna aperture of a few wavelengths.
This is in line with the near-field effect theory,
which indicates that the near-field is that part of the radiated field,
where the distance from the source,
an antenna of aperture $D$,
is shorter than the Fraunhofer distance, $d_f = 2D^2/\lambda$~\cite{NFC2011}.
As a result,
in a realistic scenario,
we would need an UDN with around $10^6$\,$\textrm{BSs/km}^{2}$ for this near-field effect to be an issue.
The red curves with square markers in Fig.~\ref{fig:theDecline} show the negative impact that
the cap on the signal power imposed by the near-field effect has on the coverage probability and the ASE.
It also shows that this only occurs at the mentioned extremely high BS densities.

\textbf{Takeaways:}
The BS density region where the near-field effect has an impact is very unlikely to be seen in practice,
as it means having one small cell BS every square meter.
This renders the near-field effect issue as negligible in practical deployments.

\section{The Challenge of the Small Cell Base Station Antenna Height\label{sec:antennaHeight}}

At the same time that researchers carried the above presented efforts to shed new light on the performance impact of the near-field effect,
the authors in~\cite{Ding2016GC_ASECrash}~\cite{8057291} discovered yet another reason why the interference power could grow faster than the signal power in a UDN:
The difference $L$ between the antenna heights of the UEs and small cell BSs.

\begin{figure}[t]
         \centering
        \includegraphics[width=0.48\textwidth]{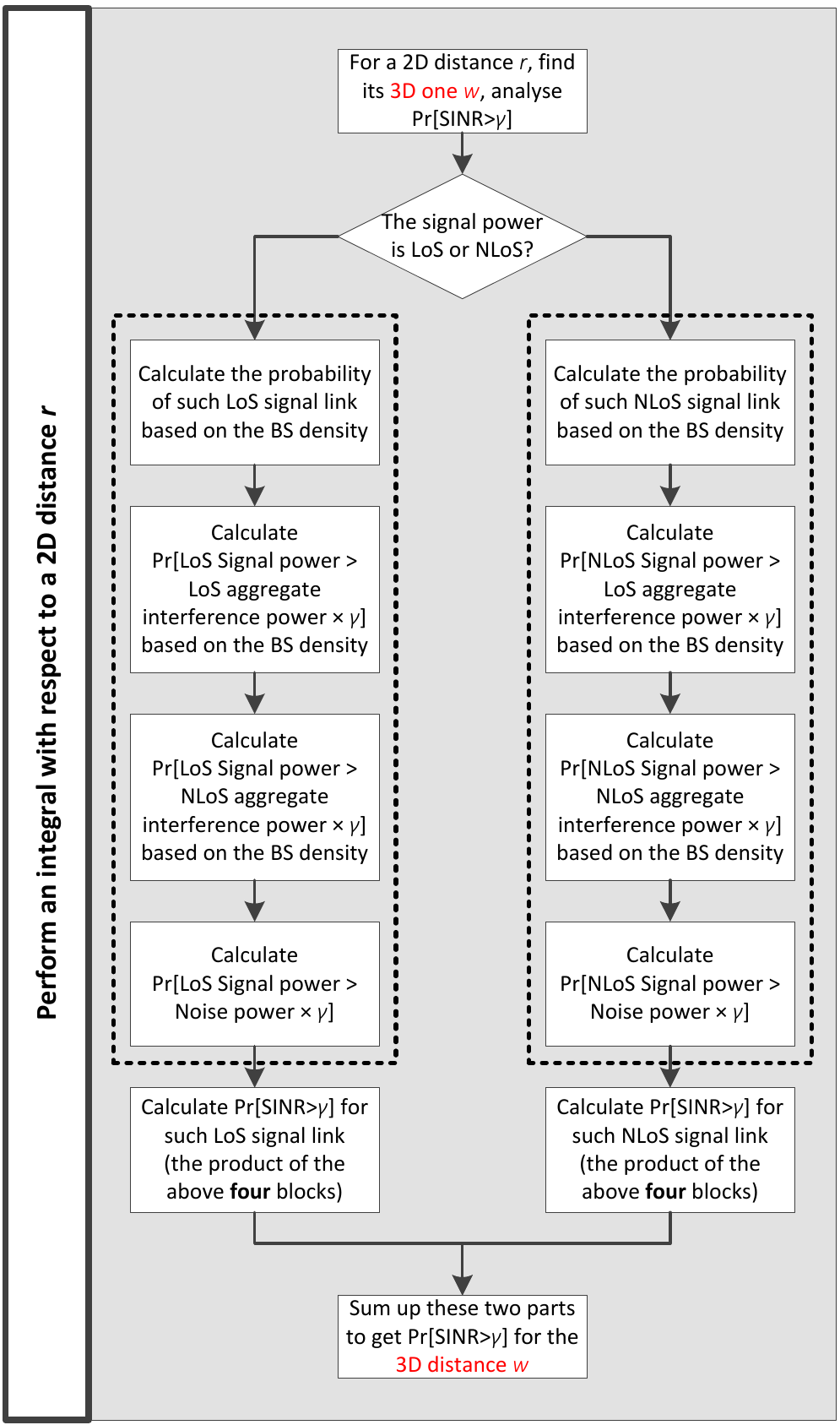}
        \caption{Logical steps within the standard stochastic geometry framework to obtain the results in~\cite{Ding2016GC_ASECrash}~\cite{8057291}, 
        considering NLoS and LoS transmissions and antenna heights, which yield the \emph{ASE Crash} phenomena.}
        \label{fig:th3}
    \end{figure}

By considering the antenna heights of both the UEs and small cell BSs in the multi-slope path loss model presented in~\cite{our_GC_paper_2015_HPPP}~\cite{our_work_TWC2016},
the authors proved that the distance between a typical UE and its interfering BSs decreases faster than the distance between such typical UE and its serving BS when densifying the network.
This is because the UE can never bridge the distance $L$,
as the UE does not climb up towards the BS!
Fig. \ref{fig:antennaHeight} depicts this lower bound on the UE-to-BS distance,
which degrades the UE's SINR,
as it poses a cap in the signal power,
while the interference power continues to increase,
since the interfering BSs continue to approach the typical UE from every direction in a densifying network.
This leads to fast declining UEs' SINRs,
and thus a potential total network outage in the UDN regime.
The pseudo code that depicts the enhancements to the stochastic geometry framework necessary to compute these new results can be found in Fig.~\ref{fig:th3}.
Compared with the logic illustrated in Fig.~\ref{fig:th2}, 
three-dimensional (3D) distances instead of 2D ones are considered in Fig.~\ref{fig:th3}.  

The green curves with triangle markers in Fig.~\ref{fig:theDecline} illustrate the results in~\cite{8057291},
and the disastrous impact that the cap on the signal power imposed by the UE and small cell BS antenna height difference has on the coverage probability and the ASE.
The latter tends to zero when the networks is ultra-dense.
We will refer to this phenomena, 
the continuous decrease in the ASE growth due to the UE and small cell BS antenna height difference, 
as \textbf{the ASE Crash} here after.

The impact of such a cap on the signal power was confirmed by~\cite{8122033} using the measurement data in~\cite{Franceschetti2004PLdata},
where the antenna height difference between the transmitter and the receiver was 4.5\,m.

It is important to note that, 
in contrast to the near-field effect,
this \emph{ASE Crash} occurs at practical small cell BS densities around $10^4$\,$\textrm{BSs/km}^{2}$ with small cell BS antenna heights of 10 meters.
This conclusion suggests that deployments of small cell BSs on street lamps or in similar locations may not be adequate,
and that it would be good to deploy small cell BSs at around human height:
not too low,
let's say above a meter from the UE antenna,
to completely avoid the near-field effect,
and not too high to mitigate the \emph{ASE Crash}.
One may argue that advanced signal processing can be used to cancel/mitigate inter-cell interference in this case,
but this would significantly increase the cost of the small cell BS,
which would be prohibitive in a UDN.
Subsequent theoretical results considering Rician and Nakagami channel models~\cite{Renzo2016intensityMatch}~\cite{8077766}~\cite{8070306}  instead of a Rayleigh one have also shown the same qualitative results,
with some quantitative deviations,
also acknowledging the \emph{ASE Crash} issue.
More recent studies considering optimal antenna downtilts have also confirmed the existence of the \emph{ASE Crash}~\cite{Yang2018downtilt}.

\textbf{Takeaways:}
Unless small cell BSs are deployed at a lower height,
close to the UE one,
the performance of an UDNs can significantly suffer.
However, it is important to mention the implications that such low small cell BS antenna deployments may have.
Among them,
it is worth highlighting the need for new small cell BS architectures,
as BSs will be at the reach of the human now,
and thus subject to tampering and other dangers in modern cities.
Small cell BSs should be so small and conformal in shape that can be hidden in plain sight to the human eye.
Moreover,
attention should be brought to new channel characteristics that may arise when having transmitters and receivers at low heights,
e.g., ground reflections, dynamic shadow fading due to fast moving object like vehicles, etc.
New research is needed in these areas.

\section{The Consequence of A Surplus of Base Stations\label{sec:IMC_and_newLaw}}

Most current mathematical analyses, 
including those presented until now, 
follow a traditional cellular paradigm,
in which the number of UEs is always much larger than the number of BSs,
and thus it is always safe to assume that there is at least one UE in the coverage area of every BS considered in the analysis.
This is also usually done for tractability reasons in the theoretical community,
but does not quite reflect the reality.
Considering that today's UE density in a urban scenario is around $\rho=$300 active\,$\textrm{UEs/km}^{2}$~\cite{Tutor_smallcell}, 
this model closely fits current sparse (50\,$\textrm{BSs/km}^{2}$) and even dense (250\,$\textrm{BSs/km}^{2}$) cellular deployments with reasonable sized cells,
but does not fit at all to UDNs ($\gg\rho$).
The reality is that some cells will be empty, with no UE, in the latter case!

In this section,
and assuming that the BS density is much larger than the UE density in a true UDN,
we show how the UDN system model was revisited to account for a finite UE density for the first time by the authors in~\cite{Ding2016IMC_GC},
and how this led to a significantly different understanding of UDNs.
Moreover, we also present and discuss the resulting more realistic capacity scaling law,
when taking this consideration into account,
which is far from a linearly increasing one.

\subsection{Exploiting A Surplus of Small Cell Base Stations\label{subsec:SurplusOfBs}}

First of all,
it is important to clarify that there are advantages in having more transmitting points than receiving ones in a network.
This has already been shown by the mMIMO technology,
where the number of antennas in a BS is much larger than the number of UEs served by such BS.
In more detail,
the large surplus of antennas in mMIMO allows to generate multiple `pencil beams', 
taking a LoS perspective for the sake of argument, 
towards different UEs,
such that these UEs can simultaneously reuse the entire spectrum managed by the cell.
Meanwhile, the interference that these UEs produce to each other, intra-cell interference, is mitigated by the narrow beams themselves (e.g., via zero-forcing precoding)~\cite{2014Marzetta}.

In a similar manner, 
a UDN can take advantage of a much larger number of small cell BSs than UEs.
In more detail,
the large surplus of small cell BSs can allow to reach the one-UE-per-cell regime,
where every UE can simultaneously reuse the entire spectrum managed by its cell,
without sharing it with other UEs.
The UEs can also benefit from a improved performance in a UDN of this nature because the small cell BSs can
\emph{i)} tune their transmit powers to the lowest possible one just to cover their small intended range,
and \emph{ii)} switch off their wireless transmissions through their idle mode capability (IMC), 
if there is no UE in their coverage areas.
These two save energy and mitigate inter-cell interference as the control signals transmitted by idle cells do not interfere neighbouring ones.
As a result of the IMC, 
it is important to note that the number of interfering cells in a UDN is at most equal to the number of active UEs,
which automatically bounds the inter-cell interference in a UDN~\cite{Ding2016IMC_GC}~\cite{8046057}.
To implement and manage this so-called IMC,
small cell BSs can take advantage of the framework initially standardised in LTE Release~13 for dynamic BS on-off~\cite{Book_SmallCells},
featuring the design of discovery reference signals for temporarily sleeping BSs.
This framework has been further enhanced in all subsequent releases, 
which substantial improvements in 5G. 
 

\subsection{The Capacity Scaling Law for Ultra-Dense Networks\label{subsec:scalingLaw}}

\begin{figure}[t]
        \centering
        \includegraphics[width=0.48\textwidth]{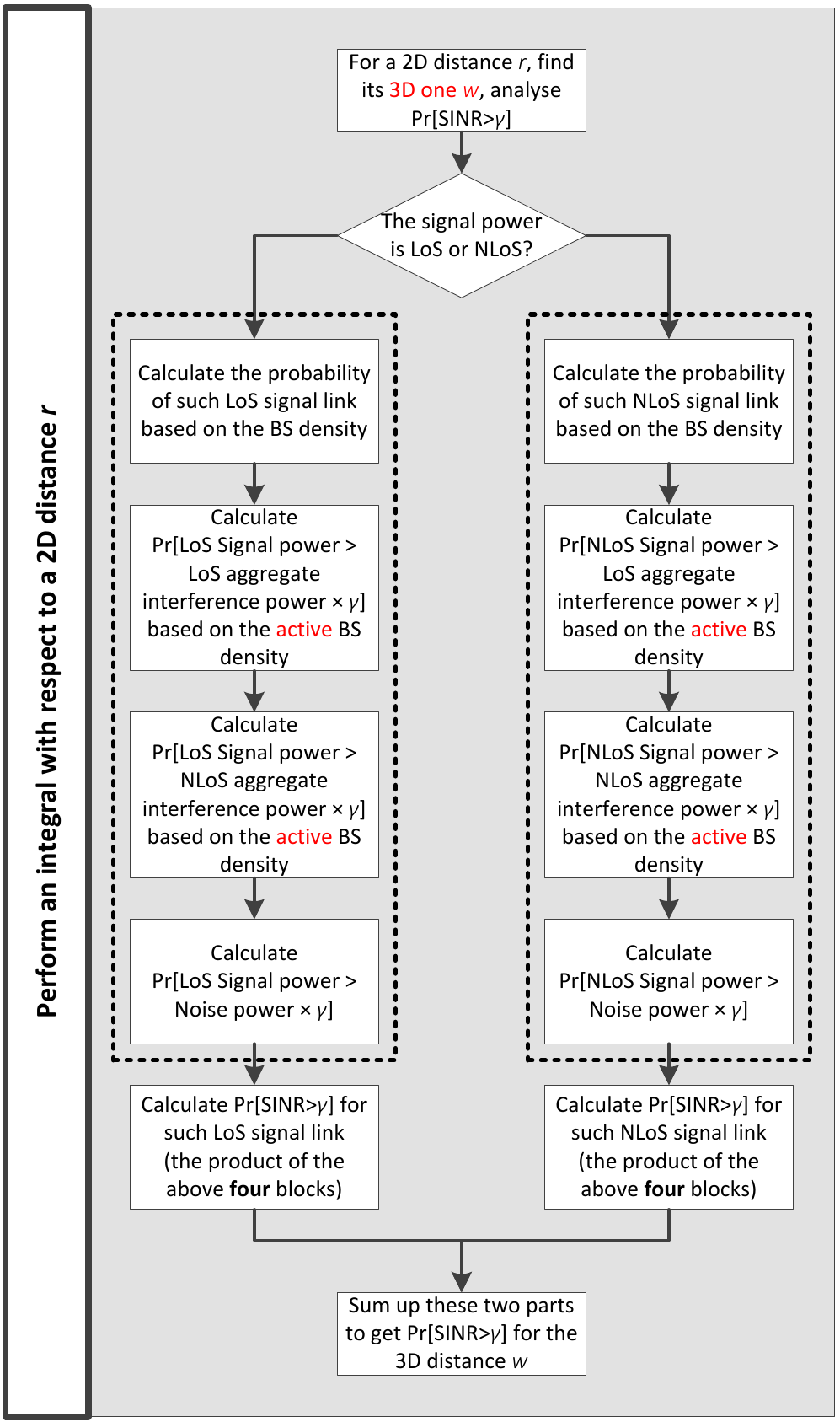}
        \caption{Logical steps within the standard stochastic geometry framework to obtain the results in~\cite{Ding2016IMC_GC}~\cite{8046057}, 
        considering NLoS and LoS trans., antenna heights and IMC, which yield the \emph{constant capacity scaling law}.}
        \label{fig:th4}
    \end{figure}

Considering the advantages of having a surplus of small cell BSs equipped with an IMC to save energy and reduce interference,  
and building on~\cite{Ding2016IMC_GC}~\cite{8046057},
in this article,
we show the more realistic capacity scaling law for UDNs, 
presented in~\cite{Ding2017capScaling}~\cite{Ding2017capScaling}, 
under the following practical and relevant system assumptions:
\emph{i)} a general multi-piece path loss model,
\emph{ii)} a non-zero small cell BS-to-UE antenna height difference,
\emph{iii)} a finite UE density, and
\emph{iv)} a surplus of small cell BSs equipped with an IMC.
In essence, this model incorporates all the modelling factors previously discussed in this article.
The pseudo code that depicts the enhancements to the stochastic geometry framework necessary to compute these new results can be found in Fig.~\ref{fig:th4}.
Compared with the logic illustrated in Fig.~\ref{fig:th3}, in this case, the interference only comes from actives BSs instead of the entire BS set. 

\begin{figure*}
    \centering
    \begin{subfigure}[b]{0.48\textwidth}
        \includegraphics[width=\textwidth]{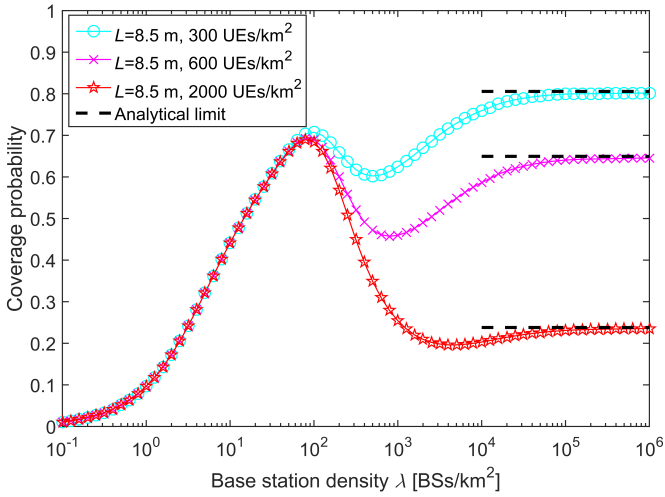}
        \caption{Coverage probability.}
        \label{fig:covProb2}
    \end{subfigure}
    \hspace{0.3cm}
    \begin{subfigure}[b]{0.48\textwidth}
        \includegraphics[width=\textwidth]{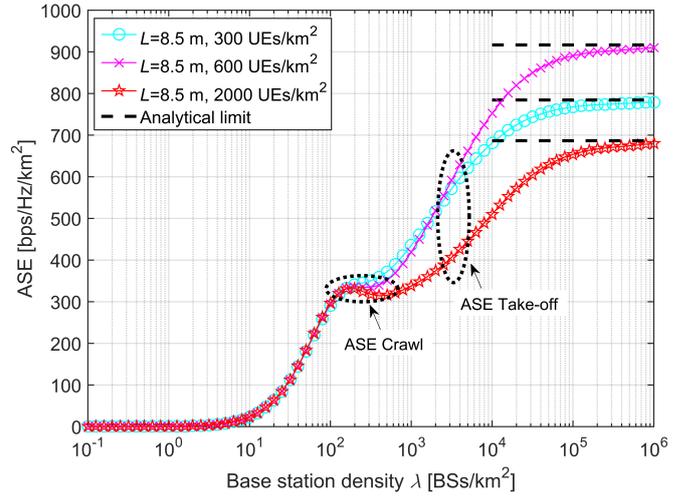}
        \caption{ASE.}
        \label{fig:ase2}
    \end{subfigure}
    \caption{The performance behaviours in coverage probability and ASE due to the surplus of BS with IMC.}\label{fig:theTakeOff}
\end{figure*}

Intuitively speaking, 
in a UDN, 
\emph{(i)} a non-zero small cell BS-to-UE antenna height difference ($L>0$) leads to a bounded signal power 
(as shown in Section~\ref{sec:antennaHeight}), and
\emph{(ii)} a finite UE density ($\rho<+\infty$) served by a surplus of small cell BSs equipped with an IMC leads to a bounded interference power 
(as discussed in Section~\ref{subsec:SurplusOfBs}).
Thus, as a consequence of a bounded signal and interference power,
the SINR becomes independent of the BS density, 
but a function of the UE density instead.  
The coverage probability becomes thus a constant when the BS density is sufficiency larger than the UE density, 
and the network then enters the UDN regime.
Such constant coverage probability leads to a constant capacity scaling law in UDNs.
The reasoning behind is the following:
\begin{itemize}
  \item each transmission link achieves a constant coverage probability performance, and
  \item the spatial reuse factor,
  i.e., the density of concurrent transmissions,
  reaches the limit of $\rho$ in UDNs.
  There cannot be more BSs active than active UEs.
\end{itemize}
The theoretical expression and proof of this more realistic capacity scaling law are relegated to~\cite{Ding2017capScaling},
as it is out of the scope of this venue,
but its results are presented in Fig.~\ref{fig:theTakeOff},
For interested readers, 
we encourage to carefully go through~\cite{Ding2017capScaling}~\cite{Ding2017capScaling}.

Fig.~\ref{fig:theTakeOff} shows two important aspects.
First, it presents the results in~\cite{8046057} and~\cite{Ding2017capScaling},
showing how both the coverage probability and the ASE increase again after the ASE Crawl (BS densities larger $10^{3}$\,$\textrm{BSs/km}^{2}$) thanks to the IMC,
as the interference power becomes bounded,
while the signal power continues to grow due to transmitter and receiver proximity. 
In this way, the ASE Crash is avoided.
We refer to this phenomena, 
as \textbf{the ASE Takeoff}.
Second, it shows how both the coverage probability and the ASE reach a constant after a given BS density,
whose specific value depends on the UE density.
The implication of this constant capacity scaling law~\cite{Ding2017capScaling} is significantly different from that of the linear capacity scaling law found in~\cite{Jeff2011}.
Specifically,
this law indicates that the network densification should be stopped at a certain level for a given UE density,
because densifying the network beyond such level is a waste of both money and energy,
since there are no obvious capacity returning gains.
This new understanding allows to develop techniques to optimise the performance of UDNs,
which are discussed in the following subsection.

\subsection{Performance Optimisation in Ultra-Dense Networks\label{subsec:performanceOptimisaton}}

UDN performance can be optimised through BS deployment and UE scheduling decisions.

The optimum BS density can be found by solving a BS deployment problem,
which aims at finding the \emph{minimum} BS density that achieves the \emph{maximum} network capacity within a performance gap of $\epsilon$-\%.
The solution to this BS deployment problem answers the fundamental question of \emph{how dense an UDN should be for a given UE density?}
This BS deployment problem can be easily solved by numerical search over the ASE results in Fig.~\ref{fig:ase2}.
For example,
for the following set of parameter values: 
$\rho=300\,\textrm{UEs/km}^{2}$, $L=8.5\,\textrm{m}$ and $\gamma_{0}=0\,\textrm{dB}$,
we can analytically find that the maximum ASE is $784.4\,\textrm{bps/Hz/km}^{2}$.
Considering a performance gap of $\epsilon=5$\,\% (i.e., a target ASE of $745.2\,\textrm{bps/Hz/km}^{2}$),
it is easy to find that the optimum BS density is $\lambda^{*}=33420\,\textrm{BSs/km}^{2}$,
for this particular case.
Any network densification beyond this level will not generate a performance gain larger than 5\,\% in terms of ASE.

\begin{figure}
    \centering
        \includegraphics[width=0.48\textwidth]{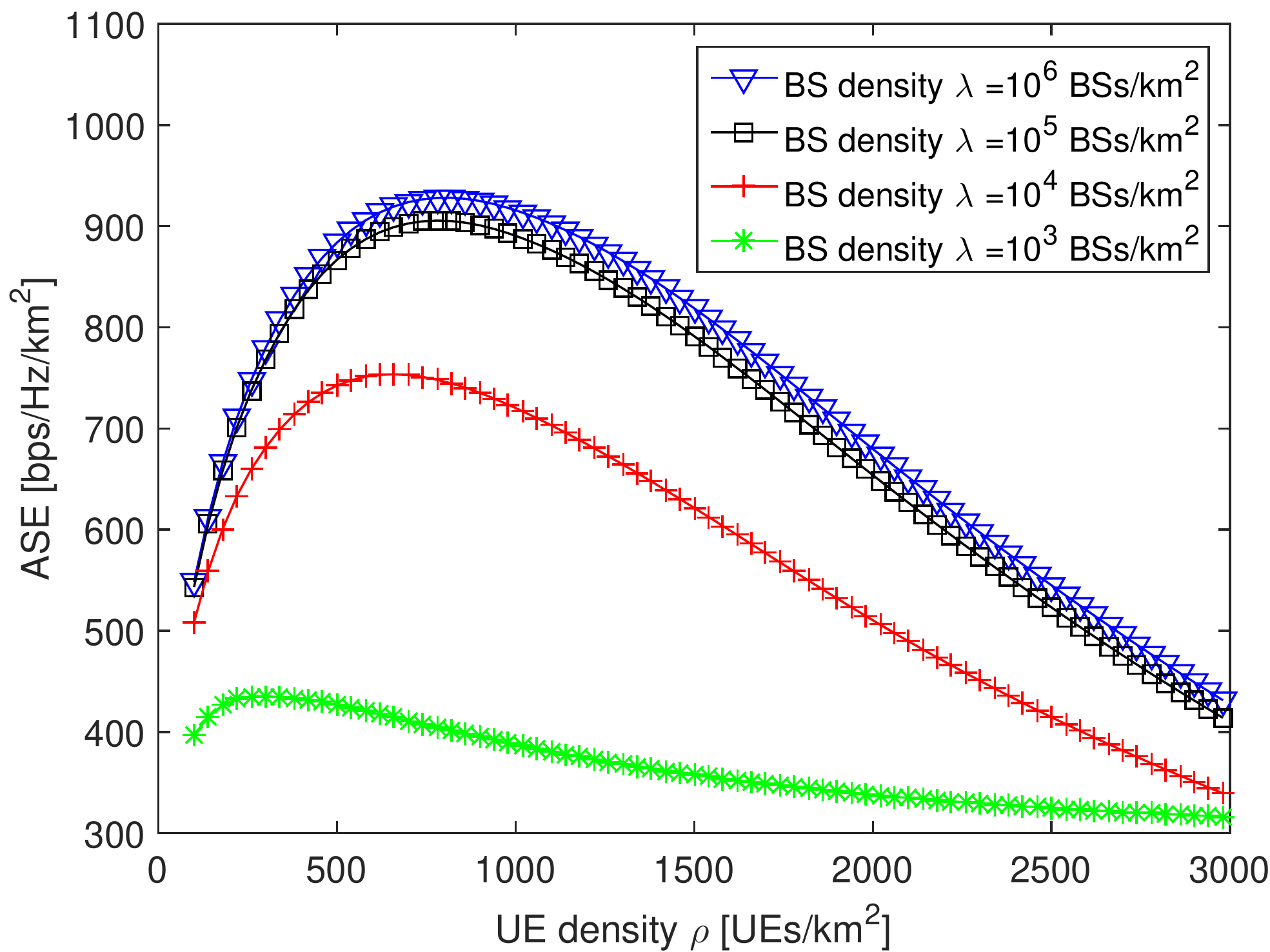}
        \caption{There is an optimal UE density that maximises ASE [bps/Hz/km$^2$] the ASE for a given BS density.}
        \label{fig:optimumUEdensity}
\end{figure}

More interestingly,
as can be derived from Fig.~\ref{fig:ase2}, 
and specifically shown in Fig.~\ref{fig:optimumUEdensity},
for a given BS density $\lambda$,
it is important to notice that the ASE performance is a concave function with respect to the UE density 
(further discussion on this topic can be found in~\cite{Ding2017capScaling}).
This is because:
\begin{itemize}
  \item
  serving more UEs requires more active BSs,
  which leads to a higher spatial reuse factor and an ASE improvement, but
  \item
  activating more BSs creates more inter-cell interference, 
  which leads to an ASE degradation.
\end{itemize}
As a result,
for a given BS density,
the ASE performance first increases and then decreases with the UE density,
which indicates the existence of an optimal active UE density.
Note that the UE density is given, 
but the active UE density can be tuned through scheduling decisions across the network. 

\begin{figure}
    \centering
        \includegraphics[width=0.48\textwidth]{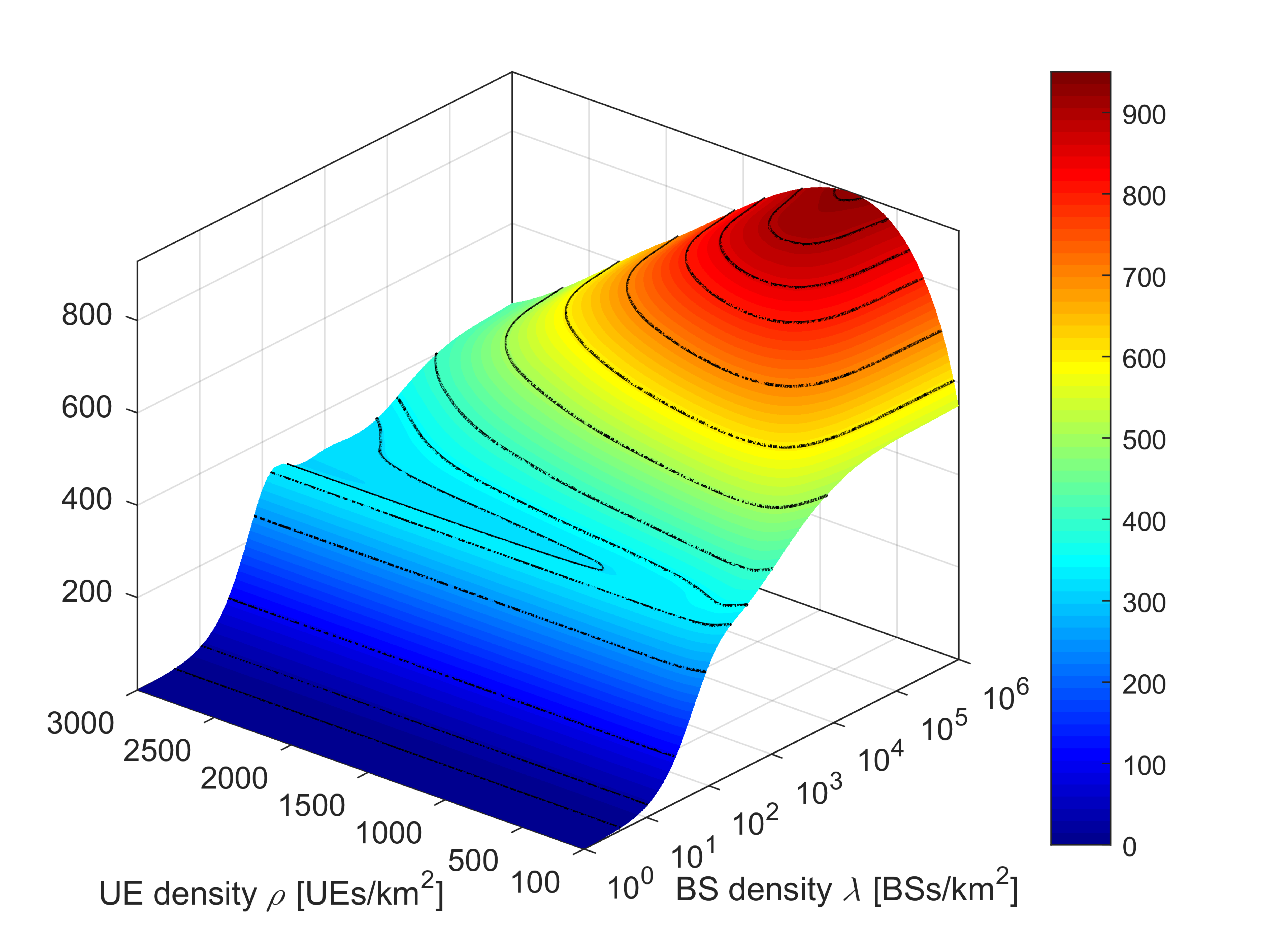}
        \caption{The ASE [bps/Hz/km$^2$] performance varies with both the BS density and the UE density.}
        \label{fig:ASE_3D}
\end{figure}

The optimal active UE density for a given BS density can be obtained by solving an active UE scheduling problem,
which aims at finding the \emph{optimal} active UE density that achieves the \emph{maximum} network capacity.
The solution to this active UE scheduling problem answers the fundamental question of \emph{what is the optimal load for a given BS density?}
This active UE scheduling problem can be  easily solved by numerical search over the ASE results in Fig.~\ref{fig:optimumUEdensity}.
For example,
for the following set of parameter values: 
$\lambda=10^{6}\,\textrm{BSs/km}^{2}$, $L=8.5\,\textrm{m}$ and $\gamma_{0}=0\,\textrm{dB}$, 
we can analytically find that the optimal active UE density is $\rho^{*}=803.6\,\textrm{UEs/km}^{2}$ with a maximum ASE of $928.2\,\textrm{bps/Hz/km}^{2}$.
Considering an alternative BS density of $\lambda=10^{4}\,\textrm{BSs/km}^{2}$,
the optimal solution is $\rho^{*}=655.4\,\textrm{UEs/km}^{2}$ with a maximum ASE of $753.6\,\textrm{bps/Hz/km}^{2}$.

As a summary, 
Fig.~\ref{fig:ASE_3D} shows how the ASE performance varies with both the BS density and the active UE density.
As can be seen from Fig.~\ref{fig:ASE_3D}, 
different combinations of the BS density and the active UE density result in different ASE results, 
and the active UE density plays a key role in achieving a satisfactory ASE performance. 

\balance
\section{Conclusion\label{sec:Conclusion}}

In this article,
we have shown that, 
from a performance stand point, 
and leaving aside deployment cost and energy consumption,
there is no dilemma on whether network densification is good or bad.
Deploying more network infrastructure,
in this case small cell BSs,
always has the potential to enhance the network and UE performance,
provided that the network is properly managed and operated. 
The right decisions are needed in terms of BS switching and UE scheduling.
We have shown that the near-field effect is not an issue for practical UDNs,
and that instead, operators should pay special attention to the channel characteristics, antenna heights,
as well as to the densities of activated BSs and served UEs per transmission time interval to avoid the performance crash due to an overwhelming interference in UDNs.
Particularly important is the finding that the coverage probability is independent of the BS density when considering an UDN and a finite density of UEs.
This indicates the existence of an optimum BS density to maximise the ASE for a given finite UE density.
For a given finite BS density,
this scaling law also indicates that there is an optimum number of UEs that can be simultaneously scheduled across the network to maximise capacity.

\section*{Acknowledgement}

For their valuable suggestions and contributions, 
for their expert guidance and the review of this article, 
we would like to thank
Prof. Guoqiang Mao,
Prof. Zihuai Lin,
Prof. Youjia Chen,
Dr. Holger Claussen, 
Dr. Lorenzo Galati-Giordano,
Dr. Giovanni Geraci,  
Dr. Adrian Garcia-Rodriguez, and
Dr. Amir Hossein Jafari. 

\section*{List of acronyms}


\hspace{-0.35cm}{\bf 2D} two-dimensional\\
{\bf 3D} three-dimensional\\
{\bf 3GPP} third generation partnership project \\
{\bf 5G} fifth generation\\
{\bf ASE} area spectral efficiency  \\
{\bf BS} base station \\
{\bf BSR} base station router \\
{\bf IMC} idle mode capability\\
{\bf LoS} line-of-sight\\
{\bf mMIMO} massive multiple-input multiple-output\\
{\bf NLoS} non-line-of-sight\\
{\bf PPP} Poisson point process \\
{\bf PGFL} probability generating functional \\
{\bf SINR} signal-to-interference-plus-noise ratio\\
{\bf SNR} signal-to-noise ratio\\
{\bf UDN} ultra-dense network \\
{\bf UE} user equipment \\
{\bf Wi-Fi} wireless fidelity

\bibliographystyle{IEEEtran}
\bibliography{Ming_library}

\end{document}